# Black Phosphorus Nanoelectromechanical Resonators Vibrating at Very High Frequencies

Zenghui Wang[1], Hao Jia[1], Xuqian Zheng[1], Rui Yang[1], Zefang Wang[2], G. J. Ye[3], X. H. Chen[3], Jie Shan[2], Philip X.-L. Feng[1,*]

[1]*Department of Electrical Engineering & Computer Science, Case School of Engineering, Case Western Reserve University, 10900 Euclid Avenue, Cleveland, OH 44106, USA*

[2]*Department of Physics, College of Arts & Sciences, Case Western Reserve University, 10900 Euclid Avenue, Cleveland, OH 44106, USA*

[3]*Hefei National Laboratory for Physical Science at Microscale and Department of Physics, University of Science and Technology of China, Hefei, Anhui 230026, China*

**We report on experimental demonstration of a new type of nanoelectromechanical resonators based on black phosphorus crystals. Facilitated by a highly efficient dry transfer technique, crystalline black phosphorus flakes are harnessed to enable drumhead resonators vibrating at high and very high frequencies (HF and VHF bands, up to ~100MHz). We investigate the resonant vibrational responses from the black phosphorus crystals by devising both electrical and optical excitation schemes, in addition to measuring the undriven thermomechanical motions in these suspended nanostructures. Flakes with thicknesses from ~200nm down to ~20nm clearly exhibit elastic characteristics transitioning from the plate to the membrane regime. Both frequency- and time-domain measurements of the nanomechanical resonances show that very thin black phosphorus crystals hold interesting promises for moveable and vibratory devices, and for semiconductor transducers where high-speed mechanical motions could be coupled to the attractive electronic and optoelectronic properties of black phosphorus.**

---

[*]Corresponding Author. Email: philip.feng@case.edu

Black phosphorus (P) is a layered material in which individual atomic layers of P are stacked together by weak van der Waals forces (similar to bulk graphite)[1]. Inside a single layer, each P atom is covalently bonded with three adjacent P atoms to form a corrugated plane of honeycomb structure (Fig. 1a, note top view of each crystal plane is in honeycomb structure). The three bonds take up all three valence electrons of P (different than graphene and graphite). This makes monolayer black P ('phosphorene') a semiconductor with a direct bandgap of ~2eV. The bandgap is reduced in few-layer phosphorene, and becomes ~0.3eV for bulk black P.[2,3,4,5,6,7,8] The bandgap and its dependence on thickness has brought mono- and few-layer phosphorene to the family of 2D crystals, especially for enabling field-effect transistors (FETs)[9,10,11,12] and optoelectronic devices with potential applications in the infrared regime,[13,14] with prototypes recently demonstrated.

In parallel to its potential for making novel electronic and optoelectronic devices, black P possesses attractive mechanical properties that are unavailable in other peer materials: it has very large strain limit (30%), and is much more stretchable (Young's modulus of $E_Y$=44GPa for single layer) than other layered materials (*e.g.*, $E_Y$=1TPa for graphene), especially in the armchair direction (*x* axis in Fig. 1a).[15] Such superior mechanical flexibility,[15,16,17] together with the exotic negative Poisson's ratio[18] arising from its corrugated atomic planes, offers unique opportunities for effectively inducing and controlling sizable strains, and thus the electronic, optoelectronic, and thermoelectric properties in this nanomaterial.[19,20,21,22,23] For example, with a 2D tension (in N/m, as in surface tension) of 0.1nN/nm, single layer $MoS_2$ can be stretched by 0.05%,[24] corresponding to a bandgap modulation of 3.7meV;[25] while a 0.45% strain can be induced in black P by the same tension (in the armchair direction),[15] which leads to a 36.4meV bandgap widening,[23] one order of magnitude greater than that in $MoS_2$. Such capability shall enable new nanoelectromechanical systems (NEMS) in which the electronic and optoelectronic properties of the nanomaterial could be efficiently tuned by the device strain,[26] thus may enable new black P devices in categories where currently only conventional materials are utilized, such as strained-channel FETs[27] and frequency-shift-based resonant infrared sensors.[28] To date, however, the exploration and implementation of black P mechanical devices have not yet been reported; such efforts have been plagued by the relative chemical activeness of black P:[8,9] it can be readily oxidized in air, and the multiple processing steps (many involving wet chemistry) required in fabricating mechanical devices from layered 2D materials (lithography, metallization, etching and suspension, *etc.*) make it particularly challenging to preserve the quality of black P crystal throughout the process. Here, we make the initial experimental study on exploring and exploiting the mechanical properties of black P to realize the first robust black P crystalline nanomechanical devices, by employing a set of specially-engineered processing and measurement techniques. We fabricate suspended black P NEMS resonators with electrical contacts using a facile dry transfer technique, minimizing sample exposure to the ambient and completely avoiding chemical processes. We characterize the material properties in vacuum, and implement nanomechanical measurements on the black P NEMS resonators using a number of experimental schemes, including Brownian motion-induced thermomechanical resonance, electrically and optically driven resonance, transient resonant motion upon pulse excitation, radio-frequency burst excitation, and resonance ring-down. Measurements in both frequency- and time-domain show that by taking proper procedures in sample preparation and measurement, black P makes robust NEMS devices exhibiting promising nanomechanical attributes.

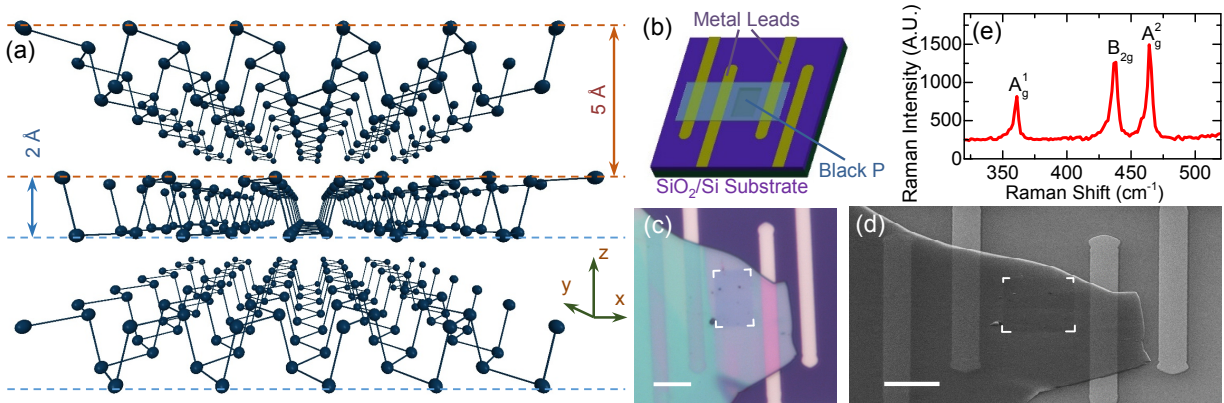

**Figure 1: Black phosphorus nanoelectromechanical device fabrication and characterization.** (a) Schematic illustration of the layered structure of black P. *x* and *y* axes are along the armchair and zigzag directions, respectively. (b) Schematic illustration of a nanomechanical resonator with an electrically contacted black P flake suspended over a rectangular microtrench. (c) Optical microscope and (d) Perspective view SEM image of a black P NEMS resonator. The scale bars are 10μm. (e) Raman spectrum measured from the device in (c) and (d), exhibiting clear black phosphorus peaks $A_g^1$, $B_{2g}$, $A_g^2$.

We fabricate suspended black P NEMS resonators using a dry transfer method. First, we synthesize black P from red P using a high temperature/pressure process (see Methods). Black P flakes are then exfoliated onto a polydimethylsiloxane (PDMS) stamp on a glass slide. After fast inspection yet careful identification of promising flakes under optical microscope, the glass slide is preserved in a vacuum chamber ($p$~5mTorr) for future transfer. We then carefully choose the pre-patterned substrate that best fits the geometry of the identified flake. The transfer is performed by aligning the desired black P flake to the target device area on the substrate (with trenches and electrodes prefabricated). We then lower the slide to bring the PDMS and substrate into contact. The PDMS is subsequently gently peeled up, and the black P flake remains on the substrate due to van der Waals forces (see Supplementary Information Fig. S1). The resulting device is immediately mounted into a vacuum chamber with optical windows and electrical feedthroughs, ready for all the subsequent optical and electrical measurements.

Our highly efficient dry transfer method is a key to successfully producing black P NEMS devices with considerably sophisticated structure (suspended device with metal contacts) while preserving the crystal quality, while traditional lithography processes and wet transfer techniques will expose the black P flakes to various wet chemical processes which can lead to undesired chemical reactions,[29,30] and require the black P crystal to remain in ambient for extended duration, significantly aggravating the undesired oxidation. With our dry transfer method, we have achieved ~70% success rate in fabricating good quality suspended black P devices.

With the device preserved in vacuum chamber, we first use Raman spectroscopy to confirm the quality of black phosphorus. Figure 1e shows the Raman spectrum of the device shown in Fig. 1c & 1d, measured *in vacuum*. The data from our multilayer black P device shows clear peeks at 361cm$^{-1}$, 438cm$^{-1}$ and 464cm$^{-1}$, corresponding to black P crystal's three dominant phonon modes, $A_g^1$, $B_{2g}$ and $A_g^2$, respectively.[1,8,10,11] Device thickness is determined by AFM after all other measurements are done (see Methods).

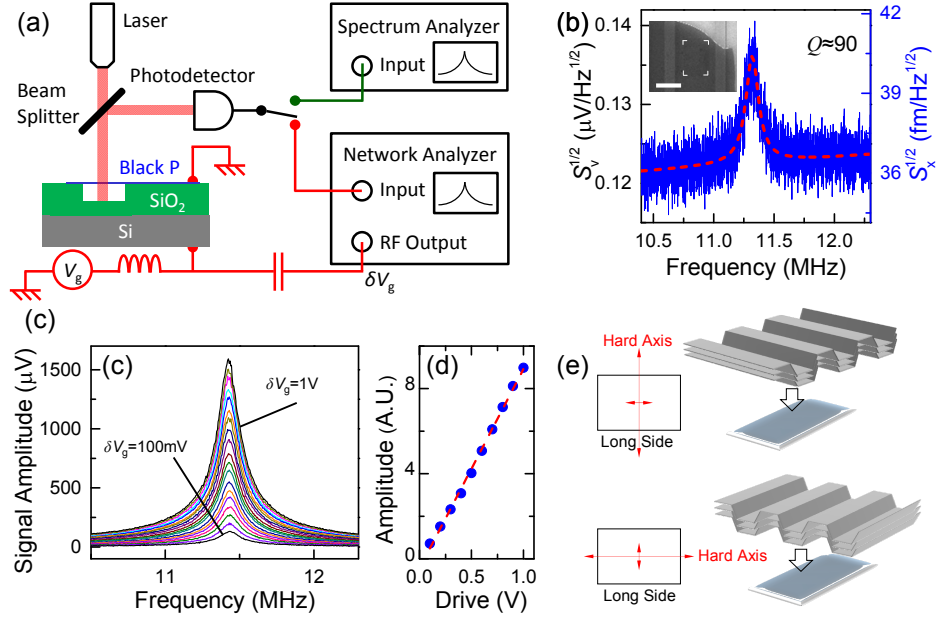

**Figure 2: Measuring frequency-domain response in black phosphorus NEMS resonator.** (a) Schematic illustration of the measurement system. The green wiring is used for measuring thermomechanical resonance, and the red wiring is the scheme for studying the driven response. (b) Measured thermomechanical resonance from the device in Fig. 1. Red dashed curve shows the fitting to a damped harmonic resonator. The right vertical axis shows the noise spectral density converted into the displacement domain. *Inset*: Top view SEM image of the device. (c) Electrically driven resonances of the same device, under different driving strength $\delta V_g$ (100mV-1V, with 50mV interval steps). (d) Measured motion amplitude *versus* driving strength, with the dashed line showing fitting to a linear function. (e) Illustration of the two alignments of the crystal axis used in the FEM simulation.

Upon successful device fabrication and material characterization, we first demonstrate high frequency black P NEMS resonators by measuring their frequency-domain resonance response, both with and without external excitation. In the absence of external driving, thermal fluctuations and dissipation processes determine the device's Brownian motion, giving rise to the thermomechanical resonance in the noise spectrum (see Supplementary Information). We measure such thermomechanical resonances using a custom-engineered ultrasensitive laser interferometry system (Fig. 2a, green wiring). Without applying any external drive, we focus a 633nm laser onto the device. Lights reflected from the different interfaces on the device structure interfere with each other, thus any out-of-plane motion in the black P flake—which modulates the optical paths—induces variations in the reflectance of the device. Such fluctuations in the reflected intensity are converted to electrical signal by a photodetector and measured with a spectrum analyzer. Figure 2b shows the fundamental-mode thermomechanical resonance of the device shown in Fig. 1. We fit the measured data to a damped harmonic resonator model (see Supplementary Information) to extract the device's resonance frequency $f_{res}$ and quality ($Q$) factor. For this 200nm-thick device, we determine $f_{res}$=11.4MHz and $Q \approx 90$. From the fitting we also obtain the displacement-to-voltage responsivity of the readout scheme; and we show the scale of the displacement-domain noise spectrum on the right vertical axis.

Through carefully engineering the optical interferometric motion-signal transduction, we achieve ~10fm/Hz$^{1/2}$-level displacement sensitivity for these devices, approaching the best interferometric displacement sensitivities enabled by SiC devices.[31]

To demonstrate electrically driven black P nanoelectromechanical resonators, we apply a voltage signal between the black P flake's electrode and the back gate, which includes a DC polarization component $V_g$ (from a DC power supply) and an AC component (output from a network analyzer, with amplitude $\delta V_g$ and frequency $\omega/2\pi$). The gate voltage generates a periodic driving force

$$F(\omega) = \frac{1}{2}\frac{dC_g}{dx}\left(V_g + \delta V_g \cos(\omega t)\right)^2 = \frac{1}{2}\frac{dC_g}{dx}\left(V_g^2 + 2V_g \delta V_g \cos(\omega t) + O(\delta V_g^2)\right) \tag{1}$$

The excited motion is detected optically by the photodetector, and the signal is measured by the same network analyzer (Fig. 2a, red wiring). Figure 2c & 2d show the resonant response of the same device under different driving strengths (from 100mV to 1V), with no nonlinearity observed. We note that for certain devices the resonance can be actuated even when the nominal DC component is seemingly at zero, $V_g$=0. This is due to the fact that the initial deviation from the charge neutrality point is providing the effective DC component in gating the device.

Employing the frequency scaling model we have developed for 2D resonators,[33] we find that this rectangular device operates in the plate regime, in which the resonance frequency is dominated by the material's elastic modulus (and insensitive to initial tension). Black phosphorus has highly anisotropic mechanical properties, with the elastic modulus in the two orthogonal in-plane directions differ by a factor of 4 to 5.[15,16,17] Therefore, for such 2D resonators the resonance frequency depends not only on the device dimension, but also on the orientation of the crystal axis (except for perfectly circular devices). Using the anisotropic Young's moduli of 37GPa/159GPa for multilayer black phosphorus,[15] we perform finite element modeling (FEM, using COMSOL) with the geometry of the 200nm device. By respectively aligning the crystalline hard (higher Young's modulus) axis with the short side and long side of the rectangle (Fig. 2e), we find that the simulated fundamental mode has $f_{res}$ of 13−18MHz, in reasonable agreement with measurement.

We further study the time-domain response of black P nanomechanical resonators. To resolve high frequency motions in *real time*, we employ an arbitrary waveform generator to drive the device and a digital oscilloscope to detect its time-resolved motions (Fig. 3a). Specifically, the generator produces a sinusoidal wave with a specified number of cycles (pulse 'train' or RF burst), which excites the black P resonator. As the pulse or burst commences, the resonator's vibration amplitude increases over time with increasing number of cycles, as energy continues to be pumped into the device. The oscillation amplitude gradually grows and saturates towards its maximum value (under the given driving strength) as the driven oscillations are fully developed (the drive from the burst balances the dissipations). Then the burst is suddenly pinched off and the periodic driving is removed, and the device motions experience a ring-down process due to un-compensated dissipations. The time-domain motional signal from the photodetector is recorded by the oscilloscope (in its channel 1, Fig. 3b top curve), and is plotted together with the driving signal (channel 2, Fig. 3b bottom curve).

The time-domain data of the ring-down process is fitted to the transient response of an undriven damped harmonic oscillator. The displacement $a$ (as a function of time $t$) is given by:[32]

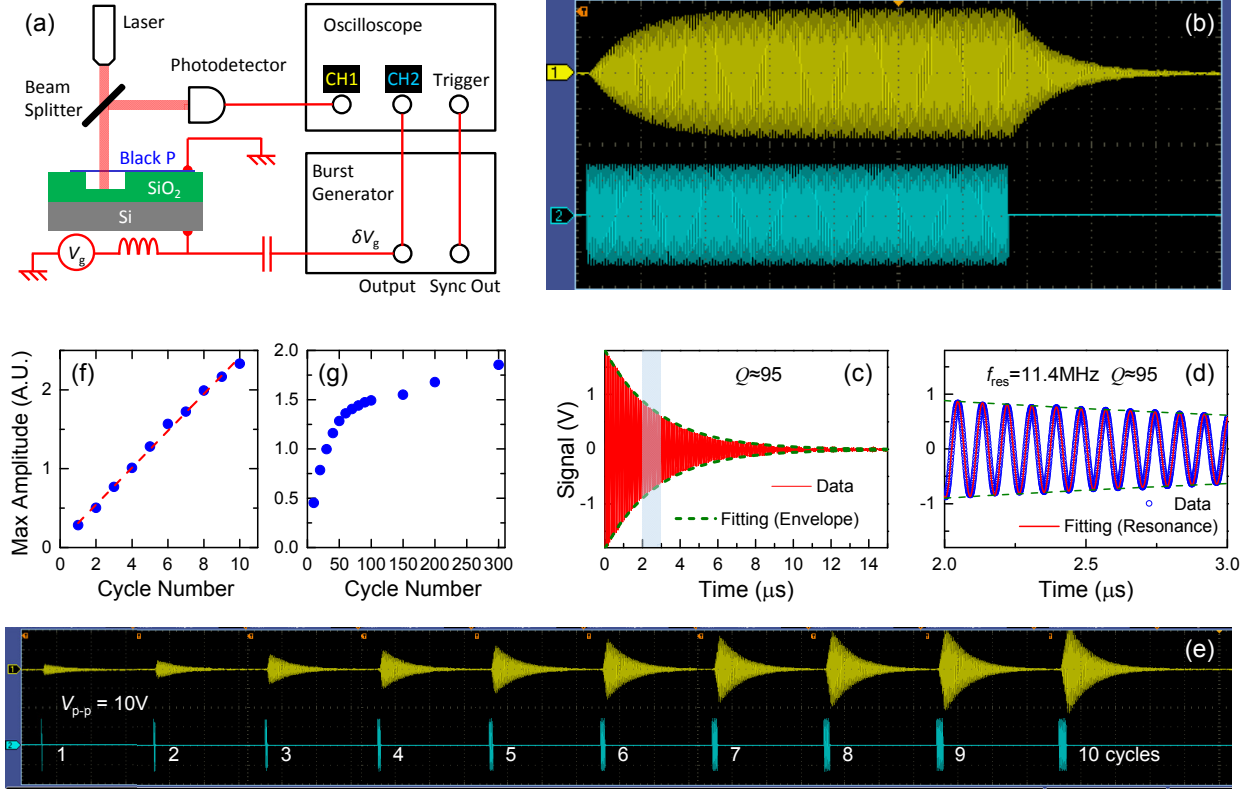

**Figure 3: Time-domain response of black phosphorus NEMS resonator.** (a) Schematic illustration of the time domain measurement system. CH1 and CH2 stands for channel 1 (yellow) and channel (cyan) of the oscilloscope (with actual screen shot shown in (b)). (b) The black P resonator's time-domain response (top yellow trace) to a 300 cycle, $V_{p-p}$=2V burst (bottom cyan trace). *Time division* (horizontal scale): 4μs. (c) Ring-down of the black P resonator. Green dashed curves show the fitting to the envelope profile. (d) Zoom-in of the shaded region in (c). Red solid line shows the fitting to the entire data set, determining both $f_{res}$ and $Q$. (e) The black P resonator's time-domain response (top yellow trace) to $V_{p-p}$=10V bursts of 1–10 cycles (bottom cyan trace). *Time division* (horizontal scale): 4μs. (f) Maximum motion amplitude as a function of burst length (number of oscillation cycles) for 1-10 cycle bursts. Red dashed line shows the fitting to a linear function. (g) Maximum motion amplitude as a function of burst length for 10–300 cycle bursts, showing amplitude saturation towards its fully developed value.

$$a(t) = A\exp\left(-\frac{t}{\tau}\right)\sin(\omega t + \varphi) \quad (2)$$

where $A$ is the initial vibration amplitude, $\omega=2\pi f_{res}$ is the angular resonance frequency, and $\varphi$ is the initial phase at $t$=0. $\tau$ is the time constant of the ring-down process, and is related to the quality factor by $Q=\pi f_{res}\tau$. Figure 3c shows the fitting of the envelope ($A\exp(-t/\tau)$) of the time-domain displacement data, which exemplifies the decaying of the motion amplitude due to dissipation. From the fitted curve we extract $Q$≈95, in very good agreement with the frequency-domain measurement. In addition, with the time-resolved measurement, we are able to directly fit Eq. 2 to the entire data set to extract both $f_{res}$ and $Q$ (Fig. 3d). The results again agree well with the frequency-domain measurement, as well as the envelope fitting (see Supplementary Information for details of the fitting).

In addition to the ring-down process, we also demonstrate time-resolved observation and calibration of the resonator's ring-up transient process. We apply short bursts (trains of sinusoidal cycles) with different numbers of cycles. The resonator rings up as energy is continuously pumped into the system, reaching maximum motion amplitude at the moment the burst ends (and undergoes ring-down immediately after). This is equivalent to the instantaneous amplitude after the same number of burst cycle for a resonator driven by a much longer burst (as in Fig. 3b). By varying the length of the short bursts, we are able to perform a "time-domain tomography" for the ring-up process. The advantage of using discrete sets of short bursts (instead of a long burst) is that very large driving amplitudes can be applied without the danger of overdriving the device into nonlinearity (as with longer bursts), which allows the fine motion with ultrasmall amplitude at the very beginning of the ring-up process be resolved and accurately characterized. Figure 3e shows the ring-up measurement with bursts of 1–10 cycles with 10V gate voltage, with the variation in maximum amplitude of the resonator clearly visible. These values are plotted in Fig. 3f as a function of cycle numbers in the bursts. From the data we clearly observe that the motion amplitude is directly proportional to the number of driving cycles, suggesting that the energy in the resonator increases parabolically at the beginning of the ring-up. As the number of driving cycles further increases (Fig. 3g), the device's vibration amplitude gradually deviates from the linear relationship and eventually saturates towards its fully developed value.

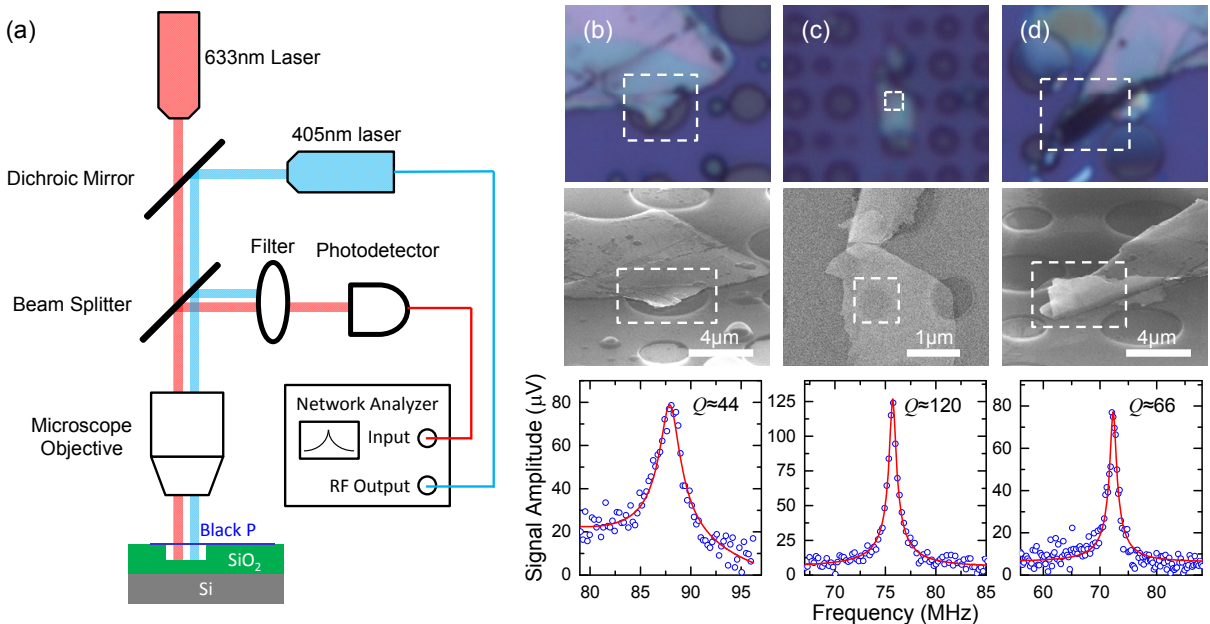

**Figure 4: Optically-driven resonances in black phosphorus resonators.** (a) Schematic of the optical excitation-detection system. (b), (c) and (d) are the optical images, SEM images, and measured mechanical resonances from 3 different black P resonators. Device in (b) is ~200nm thick and partially covers a 5μm circular trench. Device in (c) is ~190nm thick and fully covers a 0.6μm circular trench. Device in (d) has a tilted edge, forming a unique hinge-structured resonator. The boxed areas in the optical images and SEM images correspond to each other. The red curves in the data plot are fittings to a damped harmonic resonator model.

Next, we demonstrate optically driven black P nanomechanical resonators. In addition to the dry transfer methods described earlier, we also fabricate electrode-free devices by directly exfoliating black P flakes onto $SiO_2$-on-Si substrates with pre-patterned circular microtrenches. This results in suspended black P flakes either fully or partially covering the microtrenches. The fabrication process is also performed in a very fast manner to minimize the exposure to air before the resulting devices are mounted into the vacuum chamber with pressure kept at ~5mTorr.

Using a fully optical excitation/detection system,[33,34] the resonant motion of the electrode-free black P device is optothermally driven by an amplitude-modulated 405nm blue laser and interferometrically detected by a 633nm red laser.[35] As shown in Fig. 4a, the 405nm blue laser is modulated by an AC driving signal supplied by a network analyzer. We set an average on-device laser power of ~660μW and a modulation depth of 100% to achieve high motion amplitude and thus clear resonance signal. The 633nm red laser is focused on the flake surface with ~700μW on-device laser power. By sweeping the driving frequency (1-100MHz), we record the output signal using the same network analyzer, and identify nanomechanical resonance in the devices. We then fit the data to a damped harmonic resonator model, from which $f_{res}$ and $Q$ are extracted.

Figure 4b-d show 3 such electrode-free black P nanomechanical resonators with resonances in the HF and VHF radio bands. Figure 4b shows a ~200nm-thick device partially covering a 5μm-diameter circular microtrench. We measure a resonance frequency of 87.9 MHz and a $Q$ factor of ~44. Figure 4c shows a ~190nm-thick device fully covering a 0.6μm-diameter microtrench, with measured $f_{res}$=75.7MHz and $Q$≈120. Figure 4d shows a black P flake with its edge tilted up from the substrate and forms a cantilever-like hinge-structured resonator. With the highly versatile optical setup, we successfully measure resonance in this unique device, with $f_{res}$=72.3MHz and $Q$≈66. Our results show that robust nanomechanical responses can be attained in suspended black P devices with a wide range of structures and mechanical configurations.[36]

While the blue laser power is kept low to avoid excessive heating of the devices, we occasionally observe laser-induced degradation of black P flakes during optically-driven measurements. Upon extended laser irradiation, certain black P flakes appear to be affected by the focused blue laser (see Supplementary Information). Therefore, the electrical actuation scheme is potentially more advantageous in preserving black P NEMS devices, which is facilitated by the highly efficient dry transfer approach we demonstrated in this work.

Finally, we highlight the measurements of much thinner devices down to ~20nm with interesting properties. Figure 5a & b show the characteristics of a black P resonator with 22nm thickness. By fitting the measured thermomechanical resonance (Fig. 5c) to a damped harmonic resonator model, we extract $f_{res}$, $Q$, and the displacement-to-voltage responsivity. Again we achieve a displacement sensitivity at the ~10fm/$Hz^{1/2}$ level for this very thin device. In contrast to the thicker device shown in Fig. 2 (which is in the plate regime), we find that this thinner circular resonator operates in the plate-membrane transition regime, approaching the membrane limit.[33] This suggests that both the material's elastic properties and the initial tension affect the resonance frequency. Here we approximate the device as a tensioned circular disk, with Young's modulus varying between the values for the soft and hard axes of black P. With this approximation we calculate the device's $f_{res}$ range as a function of its initial tension (Fig. 5d), from which we estimate the tension to be in the range of 0.1–0.16N/m.

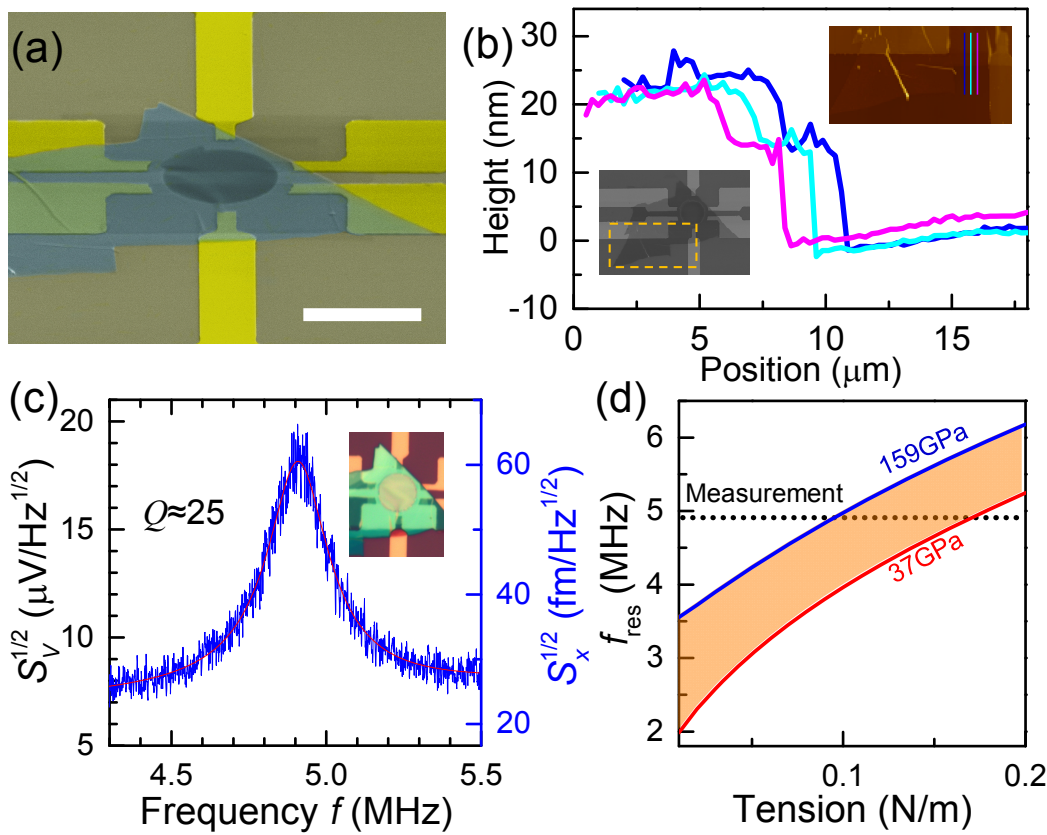

**Figure 5: Very thin black phosphorus circular membrane resonators.** (a) False-color aerial view SEM image of a 22nm-thick device. (b) Thickness profiles measured by AFM. *Top Inset*: AFM image with positions of the measured profiles indicated. *Bottom Inset*: SEM image with the AFM area indicated by the dashed line box. (c) Measured thermomechanical resonance from the device in (a). Red dashed curve shows the fitting to a damped harmonic resonator. The right vertical axis shows the noise spectral density converted into the displacement domain. *Inset*: Optical image of the device. (d) Calculated $f_{res}$ range of the device as a function of tension. The horizontal line indicates the measured frequency.

In summary, using a dry transfer technique we fabricate suspended black P devices with electrodes, while minimizing exposure of the material to the ambient. The resulting devices, with various geometries, lateral sizes, and thicknesses (from ~200nm down to ~20nm), all exhibit robust nanomechanical resonances in electrical and optical vibratory measurement schemes. The demonstrated time-domain measurement capability opens up the possibilities of real-time interrogation of coupling effects between the device's high-frequency mechanical motions with black P crystal's electronic and optoelectronic properties, by resolving the device displacement in real time. The results here show that black P holds promises for robust, new NEMS devices in which the mechanical degrees of freedom can be harnessed for enabling actuators, sensors, and dynamically-tuned electronic and optoelectronic transducers where the semiconducting and mechanical properties of black P crystals are desirable.

## METHODS

**Black Phosphorus (P) Synthesis**: Black P is synthesized under a constant pressure of 1GPa by heating red phosphorus to 1000ºC and slowly cooling to 600ºC at a cooling rate of 100ºC per hour. Red phosphorus, which was purchased from Aladdin Industrial Corporation with 99.999% metal basis, is pressed and heated in a boron nitride crucible. The high-pressure environment is provided by a cubic-anvil-type apparatus. Black P is kept in inert environment since synthesized.

**Device Fabrication**: Black P NEMS resonators are fabricated by transferring black P nanosheets onto pre-fabricated device structures. First, electrodes (5nm Cr followed by 30nm Au) are patterned onto a silicon (Si) wafer covered with 290nm thermal oxide ($SiO_2$) using photolithography followed by electron beam evaporation. Microtrenches of different sizes are then patterned with alignment to the electrodes using photolithography followed by reactive ion etch (RIE). Black P nanosheets are then transferred onto this structured substrate (see Supplementary Information).

**Interferometric Resonance Measurement**: Resonant motions of black P nanomechanical resonators are measured with a custom-built laser interferometry system. A He-Ne laser (632.8nm) is focused on the suspended black P diaphragms using a 50× microscope objective, with a spot size of ~1μm. We apply a laser power of ~700μW onto the device which assures good optical signal and does not exhibit measurable heating. Optical interferometric readout of black P device motion is accomplished by detecting the motion-modulated interference between the reflections from the black P flake-vacuum interfaces and the underneath vacuum-$SiO_2$ and $SiO_2$-Si interfaces. We have specially engineered our system to achieve ultrafine displacement sensitivities for NEMS devices, by exploiting latest advances and techniques in such schemes.[33] The vacuum chamber is maintained under moderate vacuum (~5mTorr).

**Scanning Electron Microscopy (SEM) & Atomic Force Microscopy (AFM)**: SEM images are taken inside an FEI Nova NanoLab 200 field-emission SEM, using an acceleration voltage of 30kV. AFM measurements are conducted with an Agilent N9610A AFM using tapping mode. To measure the thickness of each device, multiple traces are extracted from each scan, from which the thickness value and uncertainty are determined (see Supplementary Information).

**Acknowledgement:** We thank Jaesung Lee for useful discussions. We thank the support from Case School of Engineering, National Academy of Engineering (NAE) Grainger Foundation Frontier of Engineering (FOE) Award (FOE2013-005), CWRU Provost's ACES+ Advance Opportunity Award. Zefang Wang and Jie Shan acknowledge support from the AFOSR under grant FA9550-14-1-0268. Part of the device fabrication was performed at the Cornell NanoScale Science and Technology Facility (CNF), a member of the National Nanotechnology Infrastructure Network (NNIN), supported by the National Science Foundation (Grant ECCS-0335765).

**Electronic Supplementary Information (ESI) Available**: A supporting document with additional technical details is included as a separate PDF file.

Accepted Version of
Z. Wang, H. Jia, X. Zheng, R. Yang, Z. Wang, G.J. Ye, X.H. Chen, P.X.-L. Feng
*Nanoscale* **7**, 877-884 (2015)
DOI: 10.1039/C4NR04829F, Online Publication: Oct. 13, 2014

Accepted Version of
Z. Wang, H. Jia, X. Zheng, R. Yang, Z. Wang, G.J. Ye, X.H. Chen, P.X.-L. Feng
*Nanoscale* **7**, 877-884 (2015)
DOI: 10.1039/C4NR04829F, Online Publication: Oct. 13, 2014

# Black Phosphorus Nanoelectromechanical Resonators Vibrating at Very High Frequencies

Zenghui Wang[1], Hao Jia[1], Xuqian Zheng[1], Rui Yang[1], Zefang Wang[2], G. J. Ye[3], X. H. Chen[3], Jie Shan[2], Philip X.-L. Feng[1,*]

[1]*Department of Electrical Engineering & Computer Science, Case School of Engineering, Case Western Reserve University, 10900 Euclid Avenue, Cleveland, OH 44106, USA*

[2]*Department of Physics, College of Arts & Sciences, Case Western Reserve University, 10900 Euclid Avenue, Cleveland, OH 44106, USA*

[3]*Hefei National Laboratory for Physical Science at Microscale and Department of Physics, University of Science and Technology of China, Hefei, Anhui 230026, China*

## Table of Contents

---

[*]Corresponding Author. Email: philip.feng@case.edu

## S1. Black Phosphorus (P) Resonator Fabrication

In order to fabricate high quality suspended black P nanoelectromechanical resonators, we transfer black P flakes onto Si/SiO$_2$ substrates with pre-patterned features using a dry transfer technique specially developed and optimized for black P. Upon surveying the different existing 2D material transfer methods[1,2,3] and evaluating their respective strengths, we adopt their good features and incorporate special steps uniquely developed to facilitate fast transfer of black P, in order to minimize the material's exposure to the ambient. We first cut a small rectangular piece from PDMS film (Fig. S1a), gently peel off the protection layers on both sides (Fig. S1b), and carefully stamp it onto a clean glass slide (Fig. S1c). The black P samples are then exfoliated for dozens of times (Fig. S1d) before we press the tape onto the PDMS stamp, and gently brush the surface using a Q-tip (Fig. S1e). Black P flakes transferred onto the PDMS stamp are quickly yet carefully inspected under optical microscope to identify promising black P flakes for device fabrication, and stored in vacuum chamber immediately afterwards. By keeping the black P flakes in vacuum chamber, we have sufficient time to carefully plan for the transfer: specifically, we choose the best substrate with the most suitable pattern that fits the size and shape of the black P flake; we determine the optimal substrate mounting orientation, and we preview and examine the expected device geometry by overlaying optical images of the black P flake and the substrate in computer, in order to determine the best transfer strategy. In this way, the time duration of black P exposed to ambient environment during the transfer process is minimized, which helps prevent material degradation.

As soon as the transfer strategy is planned, we mount the pre-patterned substrate and clamp the glass slide with desired black P flake on its PDMS to our custom-built transfer stage (Fig. S1f). We move the black P flake to the target area with the micro-manipulator and align the flake with the substrate pattern under the microscope (Fig. S1g). As we gradually lower the glass slide to bring the PDMS stamp and substrate into contact (Fig. S1h), we keep fine-adjusting the position and microscope focus so that the black P flake exactly covers the desired microtrenches and electrodes. After the entire black P flake is in contact with the substrate, we gently lift up the glass slide, leaving the black P flake on the substrate by van der Waals forces (Fig. S1i). The resulting device is perfectly suspended over the microtrench and in direct contact with the metal electrodes underneath (Fig. S1j).

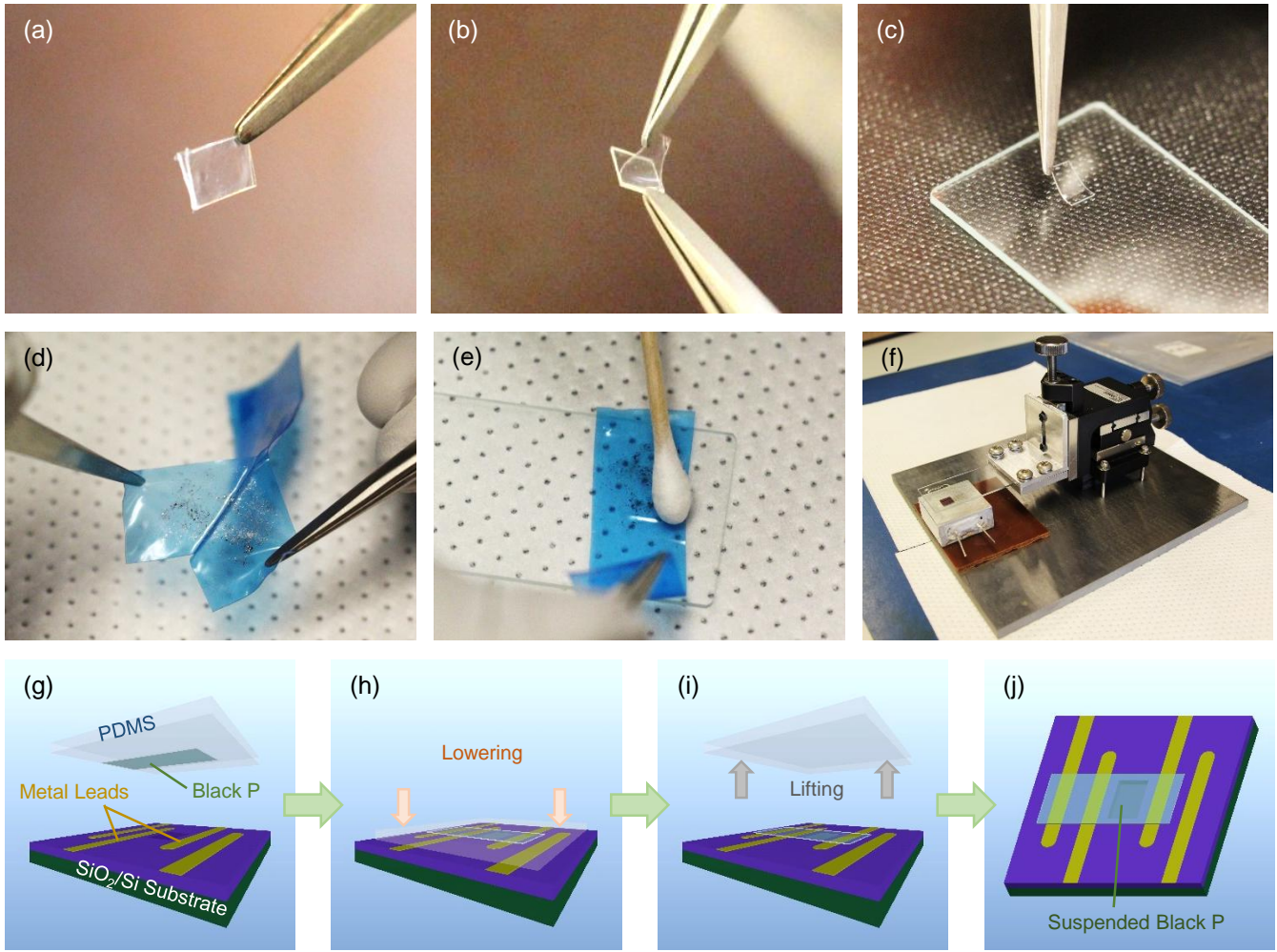

**Figure S1**: Fabrication process of suspended black P nanoelectromechanical resonators. (a-f) Images of black P flakes exfoliation and transfer preparation steps. (a) PDMS covered with protection layers on both side. (b) Peeling off the protection layers. (c) Applying the PDMS stamp onto a glass slide. (d) Pre-exfoliation of black P. (e) Final exfoliation onto PDMS stamp. (f) Transfer stage with substrate mounted and glass slide clamped. (g-j) Schematic of the transfer process. (g) Aligning the black P flake on PDMS to the prefabricated substrate patterns. (h) Lowering the PDMS stamp to bring the black P flake and substrate into contact. (i) Lifting up the PDMS stamp with the black P flake remaining on the substrate. (j) The resulting device.

## S2. Fitting the Data of Thermomechanical Resonance

The signal measured on the spectrum analyzer takes form of a resonance response on top of a frequency-dependent background (see Fig. 2b in the main text). In the frequency domain, the thermomechanical motion of a resonator is given by[4]

$$S_{x,th}^{1/2}(\omega) = \left[ \frac{4\omega_0 k_B T}{Q M_{eff}} \cdot \frac{1}{(\omega_0^2 - \omega^2)^2 + (\omega_0 \omega/Q)^2} \right]^{\frac{1}{2}}, \quad (S1)$$

and simplifies to

$$S_{x,th}^{1/2}(\omega_0) = \sqrt{\frac{4 k_B T Q}{\omega_0^3 M_{eff}}}. \quad (S2)$$

when the device is on resonance. Here, $k_B$, $T$, $\omega_0$, $Q$, and $M_{eff}$ are the Boltzmann's constant, temperature, angular resonance frequency, quality factor, and the effective mass of the device, respectively.

Assume the noise processes are uncorrelated, the total noise power spectral density (PSD) is the sum of the PSDs from individual noise processes. Thus we have $S_{v,total}^{1/2} = (S_{v,th} + S_{v,sys})^{1/2}$. Here $S_{v,th}^{1/2}$ is the thermomechanical motion noise spectral density translated into the electronic domain, through the 'displacement-to-voltage' responsivity $\Re \equiv S_{v,th}^{1/2}/S_{x,th}^{1/2}$. The other term, $S_{v,sys}^{1/2}$, is the voltage noise floor of the measurement system, which depends on the details of the detection scheme. Typically we have $S_{v,sys}^{1/2} \approx 0.1 \mu V/Hz^{1/2}$ in the vicinity of 10MHz frequency band, which slightly increases with increasing frequency. The level of $S_{v,sys}^{1/2}$ determines the off-resonance 'baseline' background ($S_{v,total}^{1/2} \approx S_{v,sys}^{1/2}$ off the resonance).

We fit our data to the expression for $S_{v,total}^{1/2}$, by using $S_{v,th}^{1/2} = \Re \times S_{x,th}^{1/2}$ and treating $S_{v,sys}^{1/2}$ as a frequency-dependent function:

$$S_{v,total}^{1/2} = (\Re^2 \times S_{x,th} + S_{v,sys})^{1/2} = \sqrt{\Re^2 \left( \frac{4\omega_0 k_B T}{Q M_{eff}} \cdot \frac{1}{(\omega_0^2 - \omega^2)^2 + (\omega_0 \omega/Q)^2} \right) + S_{v,sys}}. \quad (S3)$$

From the fitting we obtain $\omega_0$, $Q$, and $S_{v,sys}^{1/2}$ (assuming $T = 300K$). The curve in Fig. 2b of the Main Text shows a fitting to Eq. S3. From the fitting parameters we extract $\Re$, and calculate $S_x^{1/2}$ from $S_v^{1/2}$ (Main Text, Fig. 2b and Fig. 5c right vertical axis).

## S3. Fitting the Data of Time-Domain Ring-Down Measurement

In the time domain measurement, we obtain data as shown in Fig. S2a and S2c. To perform the fitting, we use the ring-down part in the curve (Fig. S2b and S2d). The $Q$ valued could be obtained from the envelope of ring-down which is expressed as an exponential term, while the high frequency oscillations of the device are described by a sinusoidal function. The complete fitting function is given by $a(t) = a_0 + A\exp\left(-\dfrac{\pi f_{res} t}{Q}\right)\sin(2\pi f_{res} t + \phi)$, where $a_0$ is the offset of the response, $A$ is the amplitude, $f_{res}$ is the resonance frequency, and $\phi$ is the initial phase. In its original form, there are five parameters one need to fit at the same time, which makes it extremely challenging. To attack the problem, we first fix $f_{res}$ and $Q$ values, which are obtained from the frequency domain data. We also estimate $A$ from the starting point amplitude of ring-down process, and $a_0$ is close to 0 in our data (as one would expect from a proper measurement scheme). Upon initializing the parameters, we first perform the fitting and obtain the values of $\phi$ and $a_0$. As the fitting converges, we unfreeze A and $Q$ and continue the fitting procedure. Using this method we obtain very good fitting to the experimental data, as shown in the insets of Fig. S2b and S2d, which show the details of small zoom-in time intervals on the entire ring-down curve. We perform fitting to ring-down measurements under bursts with different duration and strength, and obtain similar $Q$ values of 95, in good agreement with the frequency domain measurement.

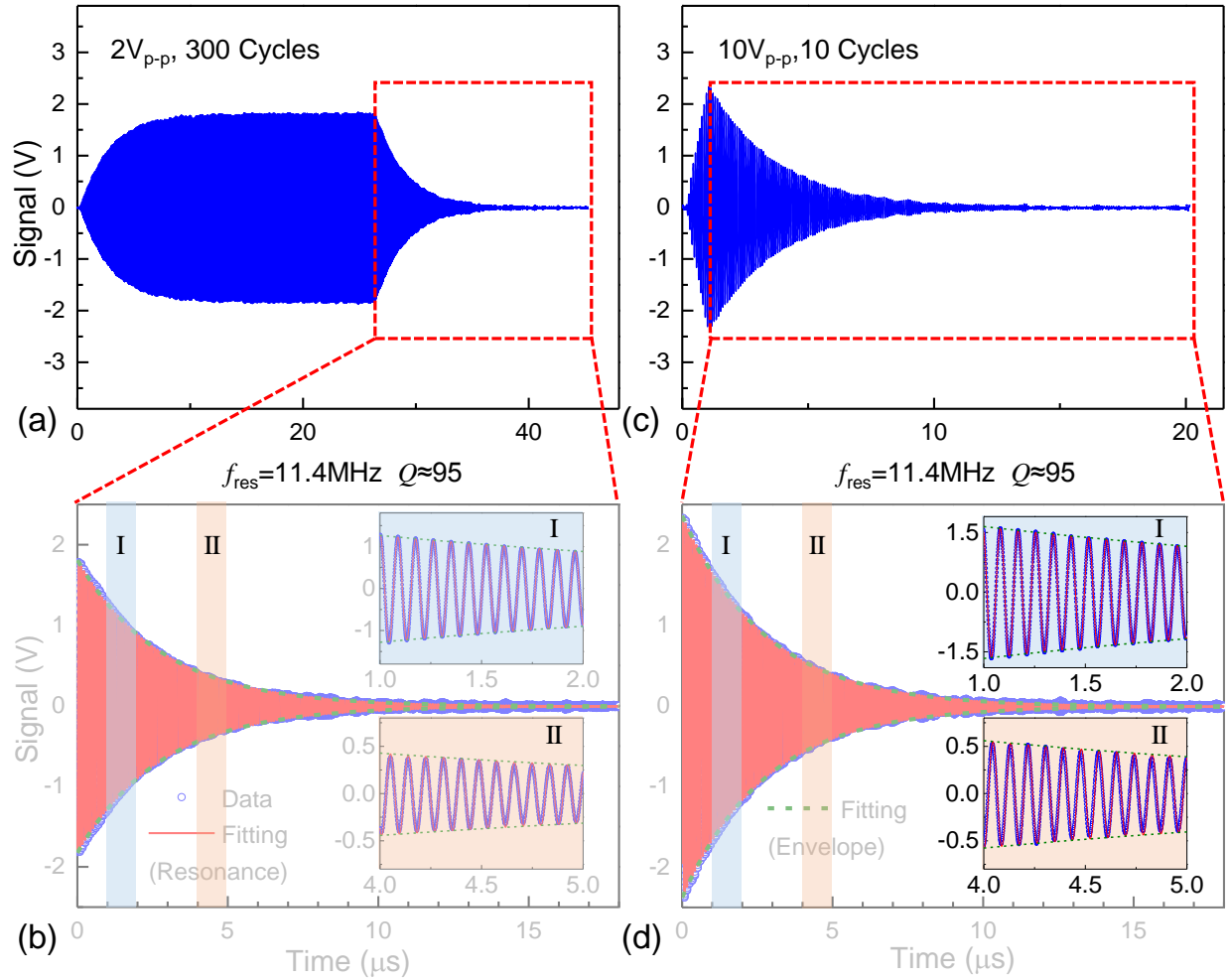

**Figure S2**: Fitting to the time-domain ring-down measurement. (a) Ring-down data for a 300 cycle burst with amplitude of $2V_{p-p}$. (b) Fitting to the ring-down data in (a) shown by the red dashed box. (c) Ring-down data for a 10 cycle burst with amplitude of $10V_{p-p}$. (d) Fitting to the ring-down data in (c) shown by the red dashed box. Insets in (b) and (d) show the zoom-in views of the fitting in the corresponding regions (I and II) as shown by the shaded area in (b) and (d).

## S4. Atomic Force Microscope (AFM) Measurement

Because the black P is easily oxidized and we have access to AFM operating in air, the thickness of the devices is measured by AFM after all the measurements are done. We then extract the step heights from the AFM data, as shown in Fig. S3.

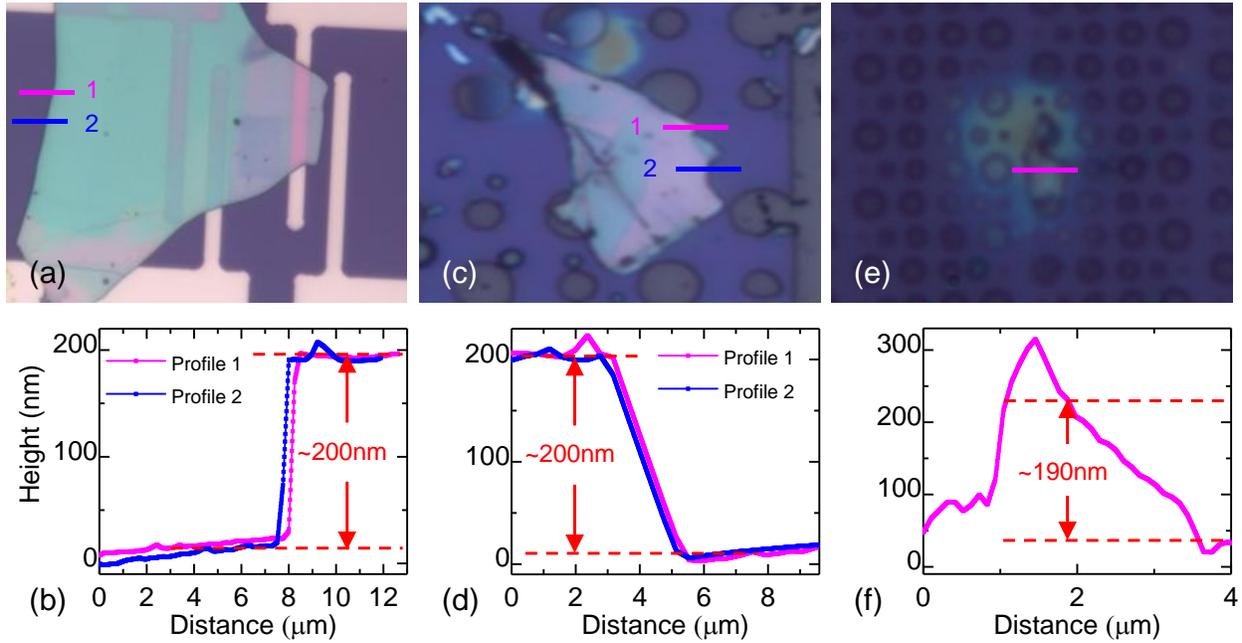

**Figure S3**: Device thickness measurement with AFM. (a,c,e) Optical microscope images of the devices. (b,d,f) Height measurement of the solid lines indicated in (a), (c), and (e), respectively. (a,b) show the same device in Fig. 1 of Main Text, (c,d) show the device in Fig. 4b and 4d of Main Text, and (e,f) show the device in Fig. 4c of Main Text.

## S5. Laser-Induced Degradation of Black P

We observe laser-induced degradation ("laser burning") of black P in optically-driven measurements.

The suspended black P resonator is driven by a modulated 405nm blue laser. We set the DC bias of the laser modulator at 30%, corresponding to an output power of 0.66mW into the microscope, and achieve 100% modulation (amplitude of AC signal equals to DC level) by adjusting the modulation depth and driving signal amplitude. The resonance of the suspended black P resonator is detected interferometrically with a 633nm red laser focused on the surface with <0.7mW on-device power.

Figure S4 shows the "laser burning" process of a black P flake. This black P flake has a thickness of ~200nm, fully covers a 0.6μm hole in the middle, and partially covers 0.8μm hole on the right and a 1.6μm hole on the bottom (as shown in Fig. 4c in the Main Text). Initial blue and red laser positions are illustrated in Fig. S4a. The blue laser is used to photothermally drive the device, and the red laser is focused at the center of the fully-covered circular area. Figure S4b and c show the obvious "burning" effect when lasers are kept at the initial positions during the measurement. We can clearly observe the expansion of "burning" area induced by the blue laser heating effect. We then change the red laser position measure the different devices on this flake (Fig. S4d-f). The blue laser moves accordingly. We clearly observe additional expansion of the "burning" area. Similar effect has been observed in other devices (Fig. 4d in the Main Text). We note that this is the first time that such laser-induced degradation in black P is observed and reported. While additional studies are needed to fully understand and quantify this effect, our results show that laser power should be carefully controlled to limit the heating effect during optical measurements of black P.

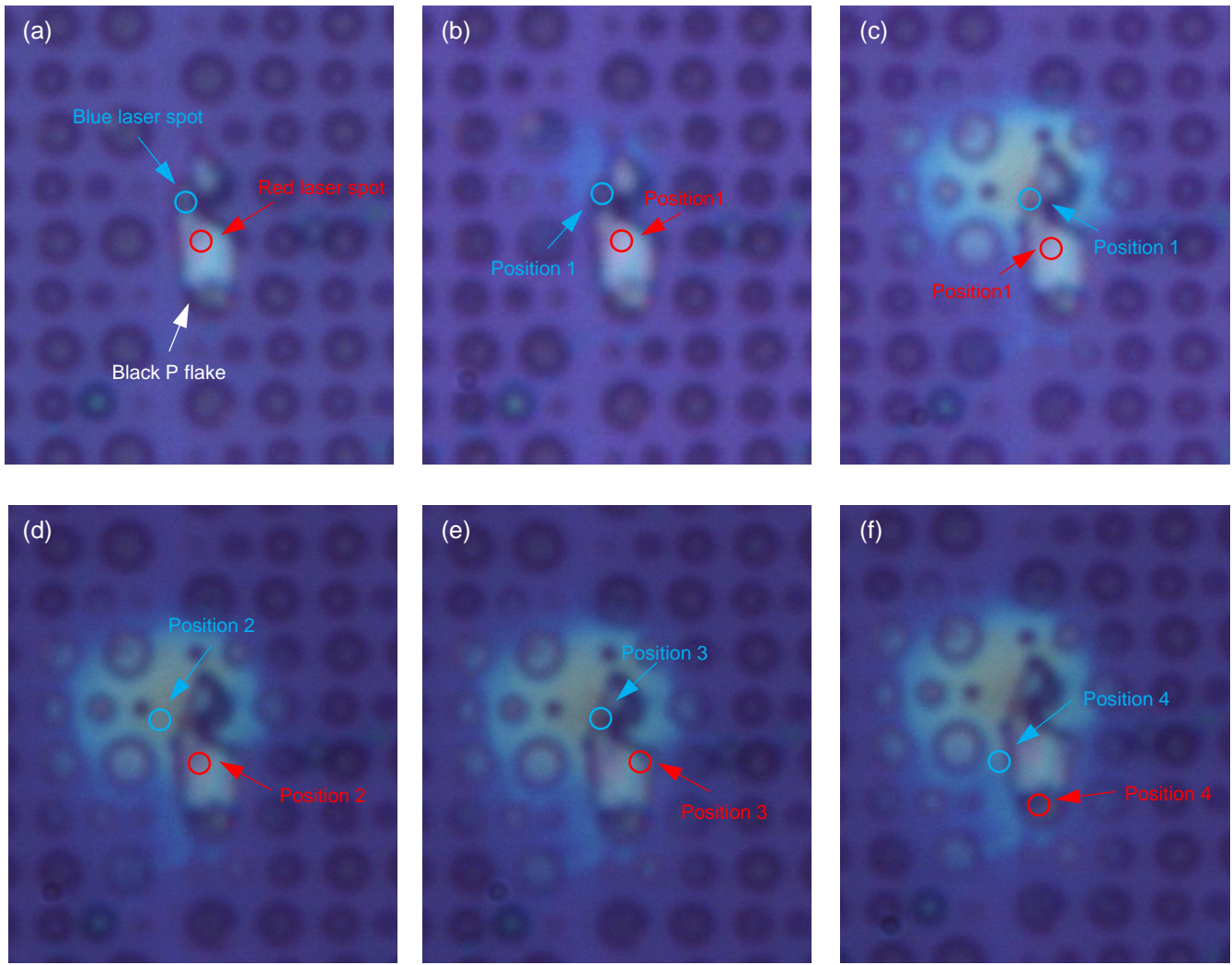

Fig. S4. Laser-induced black P degradation during optically-driven resonance measurement. (a) A black P flake fully covering a 0.6μm circular trench in the middle and partially covering a 0.8μm circular trench on the right and a 1.6μm trench on the bottom. Initial positions of the driving (blue) and detecting (red) lasers are indicated by the blue and red circles. (b) and (c) show increased "burning" by continuous heating of blue laser at the initial position during the measurement. (d), (e) and (f) show further spreading of "burning" area along with the blue laser when lasers are moved to different positions during the measurement.